\documentclass[a4paper,12pt]{article}

\usepackage{a4}
\usepackage{comment}
\usepackage{graphics}
\usepackage{graphicx}
\usepackage{latexsym}
\usepackage{amsfonts}
\usepackage{amssymb}
\usepackage{amsmath}
\usepackage{slashed}
\usepackage{feynmp}
\usepackage{hyperref}
\usepackage{url}
\usepackage{color}
\usepackage{cancel}
\usepackage{multirow,array}
\usepackage{dcolumn}
\usepackage{bm}
\usepackage{mathtools}
\usepackage[linesnumbered,ruled,vlined]{algorithm2e}
\SetAlFnt{\footnotesize}
\SetAlCapFnt{\small}
\SetAlCapNameFnt{\small}
\usepackage{authblk}
\usepackage{etoolbox}
\usepackage{float}
\usepackage{array}
\usepackage{longtable}
\usepackage{pdflscape}
\usepackage{booktabs}
\usepackage{multirow}
\usepackage{mathtools} 
\usepackage{braket}
\usepackage{orcidlink}

\usepackage{xcolor}
\usepackage{minted}
\usepackage{float}

\setminted{
  linenos,
  breaklines,
  breakanywhere,
  fontsize=\scriptsize,
  frame=single,
  framesep=2mm,
  baselinestretch=1.0,
  samepage,
  stepnumber=1,
  highlightcolor=yellow
}

\title{
  Generative quantum eigensolver with constrained circuit-cutting overhead
}

\author{Junya Nakamura\thanks{junya.nakamura@pwc.com} \orcidlink{0000-0001-9504-1371}}
\author{Shinichiro Sanji}
\affil{Technology Laboratory, PwC Consulting LLC, Tokyo, Japan}

\date{}

\begin{document}

\maketitle

\begin{abstract}

Generative quantum eigensolver (GQE) is a hybrid quantum-classical algorithm that iteratively trains a classical generative machine learning model such that the model can generate quantum circuits with desired properties such as approximating molecular ground states.
It offers as many potential applications and as much flexibility as variational quantum eigensolvers, while avoiding the problem of barren plateaus.
Quantum circuit cutting (QCC) is a technique to perform quantum computations that require more qubits than available on single quantum devices.
It comes with considerable sampling overhead depending on the structure of the circuit to be cut and how the circuit is cut. To make QCC practical, therefore, the circuits to be cut must be designed such that their execution is meaningful and QCC overhead is kept small.
In this work, we extend GQE such that the generative model only produces circuits whose overhead by QCC is upper-bounded, while retaining the original purpose of GQE.
Consequently, our proposal not only enhances the applicability of GQE through the use of QCC, but also provides a practical application for QCC.
Using a transformer decoder implementation of GQE, we evaluate our method through simulated ground state search experiments on the $\mathrm{BeH_2^{}}$ molecule.
A new loss function and a hybrid online/offline training strategy are also introduced and it is observed that these tools improve convergence and final energy values.

\end{abstract}
\newpage
\section{Introduction}

Variational quantum algorithms (VQAs) are hybrid quantum-classical algorithms that gradually optimizes a parametrized quantum circuit with classical computers by monitoring the outputs of the quantum circuit with quantum computers~\cite{cerezo2021variational}. VQAs can have a broad range of applications by choosing an appropriate objective function to optimize. Representative examples include the variational quantum eigensolver (VQE) for ground state searches of chemical systems~\cite{yung2014transistor, peruzzo2014variational}, the quantum approximate optimization algorithm (QAOA) for solving combinatorial optimization problems~\cite{farhi2014quantum}, and quantum machine learning for supervised learning~\cite{mitarai2018quantum,Havl_ek_2019,farhi2018classificationquantumneuralnetworks,Schuld_2020}. Quantum circuits used in VQAs have high flexibility with respect to several requirements from quantum computing devices such as coherence time and gate sets, which makes them attractive for their implementations on current noisy intermediate-scale quantum (NISQ)~\cite{preskill2018quantum} or future early fault-tolerant quantum computing (early-FTQC)~\cite{suzuki2022quantum,Lin_2022,Akahoshi_2024} devices. However, the problem of barren plateaus that the gradient of the objective functions becomes exponentially small as a function of the number of qubits and must be solved for achieving quantum advantage has been reported~\cite{mcclean2018barren, wang2021noise}.

Inspired by the remarkable success of large language models (LLMs) and their underlying transformer architectures in a wide range of domains (e.g.,~\cite{vaswani2017attention, radford2019language, dosovitskiy2021imageworth16x16words, zhao2021pointtransformer}), the application of LLMs to quantum circuit generation has become an active area of research in recent years~\cite{nakaji2024generative, daimon2024quantum, minami2025generative, tyagin2025qaoa, ueda2025optimizing, jern2025fine}.
Among these studies, the generative quantum eigensolver (GQE)~\cite{nakaji2024generative} has been proposed as a method that offers as many potential applications and as much flexibility as VQAs, while avoiding the problem of barren plateaus.
The GQE is also a hybrid quantum-classical algorithm that iteratively trains a classical generative machine learning model by monitoring the outputs of the quantum circuits that the model produces, until the model can generate the desired circuit.

Quantum circuit cutting (QCC) has been proposed as a technique to perform quantum computations that require more qubits than available on single quantum devices~\cite{Peng_2020, Mitarai_2021}. This technique partitions a large quantum circuit into subcircuits that are smaller in terms of the number of qubits and gates, samples them independently on quantum devices, and classically post-processes the outcomes to reconstruct the output of the original circuit.
While QCC offers the attractive advantage of executing only subcircuits of small size, it comes with considerable sampling overhead, which appears as the number of circuit runs (often called shots) that is required to achieve a desired accuracy of the final output. The QCC overhead depends on the circuit to be cut and how the circuit is cut.
To make QCC practical, the circuits to be cut must be designed such that their execution is meaningful and the QCC overhead is kept small.

In this work, we extend GQE such that the classical generative model produces only circuits whose QCC overhead is upper-bounded. The original purpose of GQE, i.e. generating quantum circuits with desired properties such as describing molecular ground states, is retained.
Consequently, our proposal not only enhances the applicability of GQE through the use of QCC, but also provides a practical application for QCC.
We also introduce a loss function and hybrid approach as two new methods to improve the training of GQE.
We implement GQE with a transformer decoder and numerically validate our methods through the ground state search experiments on the $\mathrm{BeH_2^{}}$ molecule.

The main contributions of this work are as follows.
\begin{itemize}
  \item We propose an efficient constrained circuit generation scheme for GQE, in which the generated circuits are guaranteed to satisfy the upper-bound constraint on the QCC overhead.
  \item We propose a loss-masking procedure that makes the training objective consistent with the constrained circuit generation process.
  \item We propose a new loss function and a hybrid online/offline training strategy, and demonstrate their effectiveness in simulated ground state search experiments on the $\mathrm{BeH_2^{}}$ molecule.
\end{itemize}

This paper is organized as follows.
In Sec.~\ref{sec:gqe_qcc}, we describe the approach for the circuit generation in GQE under constraints on the QCC overhead.
In Sec.~\ref{sec:pdpo_hbd},  we introduce a new loss function for GQE and the hybrid approach that combines online and offline training of GQE.
In Sec.~\ref{sec:numerical}, we describe our experimental setup and present numerical results. We conclude in Sec.~\ref{sec:conclusion}.

\section{Generative quantum eigensolver with constrained sampling overhead for quantum circuit cutting}\label{sec:gqe_qcc}

In this section, we first briefly review Generative quantum eigensolver (GQE) and Quantum circuit cutting (QCC) in \ref{sec:gqe} and \ref{seq:qcc}, respectively, focusing on topics relevant to our study. Then, in \ref{sec:gen-cut}, we explain our approach to extend GQE such that the generative model produces only circuits whose QCC overhead is upper-bounded.

\subsection{Generative quantum eigensolver}\label{sec:gqe}
In GQE~\cite{nakaji2024generative}, we prepare in advance an operator pool $G$ consisting of $L$ operators, where the operator pool and each operator are analogous to the vocabulary and a token in LLMs, respectively.
The operators in $G$ are arbitrary. In ref.~\cite{nakaji2024generative}, the unitary coupled-cluster singles and doubles (UCCSD) excitations ans\"atze derived from the target molecule  are employed with a set of different small angle parameters. In ref.~\cite{minami2025generative}, basic one- and two-qubit gates such as Hadamard gates, rotation gates and CNOT gates are employed.
Our choice for operators in $G$ is the VQE ans\"atze that have been proposed in ref.~\cite{Anselmetti_2021} and are implemented in PennyLane~\cite{bergholm2022pennylaneautomaticdifferentiationhybrid} as \texttt{qml.DoubleExcitation} and \texttt{qml.SingleExcitation}.

The choice of the classical generative model in GQE is arbitrary. Following ref.~\cite{nakaji2024generative}, we employ a transformer decoder based on the GPT-2 architecture~\cite{vaswani2017attention, radford2019language}.
The model receives as input the sequence of operator IDs generated so far and outputs logits for  the operators in $G$. Here the logits are unnormalized probabilities, from which the next operator probabilities are obtained by applying the softmax function.
The next operator ID is sampled according to these probabilities and appended to the sequence.
For instance, assume that the sequence of operator IDs $\vec{j} = [0, j_1^{}, j_2^{}, \cdots, j_{t-1}^{}]$ has been generated up to step $t-1$, where each $j_i^{}$ is an integer between $1$ and $L$ that identifies one of the operators in $G$ (i.e. $j_i^{} \in \{1, 2, \cdots, L\}$).
At step $t$, the model receives $\vec{j}$ as input and  outputs logits for the operators in $G$. Then, the next operator ID $j_t^{}$ is sampled according to the probabilities obtained from these logits and appended to the sequence, yielding $\vec{j} = [0, j_1^{}, j_2^{}, \cdots, j_{t-1}^{}, j_t^{}]$.
This process is initialized with $\vec{j} = [0]$ and repeated until $j_N^{}$ is generated, where $N$ is the number of operators to be generated and specified in advance.
Note that the operator ID $0$ is reserved for initial state specification, therefore it is not included in the pool $G$ and is not generated by the model.

We can construct a quantum circuit by sequentially applying the operators specified by $\vec{j}$ on a state initialized to a specific state such as the all-zero state (i.e. $\ket{0}^{\otimes n}$ state for a $n$-qubit circuit) or the Hartree-Fock state for chemical applications. Therefore, one sequence $\vec{j}$ uniquely determines one quantum circuit.
We denote by $E(\vec{j})$ the expectation value of a target Hamiltonian computed by running the circuit identified by $\vec{j}$.
GQE has two phases, namely the pre-training phase and the training phase. These can be regarded as offline and online learning, respectively. During the pre-training phase, a model is trained from scratch on a static dataset ${\cal D} = \bigl\{(\vec{j}^{(i)}, E(\vec{j}^{(i)})) \bigr\}_{i=1}^{M_D}$ which is prepared in advance.
During the training phase, the model is initialized either from the pre-trained model or from scratch, and is trained using a dataset generated online by the model being trained at each weight update step.

We denote by $p_{\theta}^{}(\vec{j})$  the probability distribution of $\vec{j}$ induced by the generative model, where $\theta$ denotes the model weights.
Given the purpose of GQE, we want to update the model weights so that the probability $p_{\theta}^{}(\vec{j})$ is maximized for $\vec{j}$ with the lowest value of $E(\vec{j})$.
When using backpropagation to update the model weights, the loss function must be constructed from differentiable model outputs, such as logits and $p_{\theta}^{}(\vec{j})$. Therefore, $E(\vec{j})$  cannot be used directly as the loss function.
In ref.~\cite{nakaji2024generative}, {\it logit-matching} is proposed as the loss function, where the loss function is minimized when the sum of logits matches with $E(\vec{j})$. This choice makes the model generate $\vec{j}$ following a Boltzmann distribution, so that  $\vec{j}$ with lower values of $E(\vec{j})$ are generated with higher probability.
In ref.~\cite{minami2025generative},  direct preference optimization (DPO)~\cite{rafailov2024directpreferenceoptimizationlanguage, xu2024contrastivepreferenceoptimizationpushing} is employed as the loss function. We also use DPO and its variant, and explain these in Sec.~\ref{sec:pdpo}.

\subsection{Quantum circuit cutting}\label{seq:qcc}

Quantum circuit cutting (QCC) can be in general performed as a quasi-probabilistic simulation of quantum channels, i.e. if a quantum channel $\Phi$ can be decomposed as $\Phi=\sum_i c_i^{} \Phi_i^{}$ where $c_i^{}$ and $\Phi_i^{}$ are a complex coefficient and a quantum channel, respectively,  then $\Phi$ can be simulated in a Monte Carlo manner by sampling $\Phi_i^{}$ with probability proportional to  $|c_i^{}|$ and classically post-processing the phase of $c_i^{}$~\cite{mitarai2021overhead}.
The sampling overhead for simulating $\Phi$ using this decomposition is quantified by $\sum_i |c_i^{}|$.
QCC is categorized into {\it time-like} cut~\cite{Peng_2020}, where the identity channel is decomposed into measurements and state-preparation channels, and {\it space-like} cut~\cite{Mitarai_2021}, where a non-local unitary channel is decomposed into local channels. Here we focus on {\it space-like} cuts.
We denote by $\mathcal{O}$ the quantum channel induced by the operator $O$, i.e.
$\mathcal{O}(\cdot)=O(\cdot)O^{\dagger}$.
As an example of {\it space-like} cuts, one possible quasi-probabilistic decomposition for the two-qubit Z rotation gate, $R_{Z_i^{}Z_j^{}}^{}(\theta)=e^{-i\frac{\theta}{2}Z_i^{} \otimes Z_j^{}}$, between qubit $i$ and qubit $j$ is given by~\cite{Mitarai_2021},
\begin{align}
  \mathcal{R}_{Z_i^{}Z_j^{}}^{}(\theta) = & \cos^2_{}{(\theta/2)} \mathcal{I}_i^{} \otimes \mathcal{I}_j^{} + \sin^2_{}{(\theta/2)} \mathcal{Z}_i^{} \otimes \mathcal{Z}_j^{} \nonumber \\  &- \cos{(\theta/2)} \sin{(\theta/2)} \sum_{(\alpha_1^{}, \alpha_2^{})\in \{\pm 1\}^2_{}}\alpha_1^{}\alpha_2^{} \Bigl[ \mathcal{M}_{Z_i^{}}^{}(\alpha_1^{}) \otimes \mathcal{R}_{Z_j^{}}(-\alpha_2^{}\pi/2)  \nonumber \\
   &+ \mathcal{R}_{Z_i^{}}(-\alpha_1^{}\pi/2) \otimes \mathcal{M}_{Z_j^{}}^{}(\alpha_2^{})
  \Bigr],\label{eq:rzz_mf}
\end{align}
where  $M_{Z}^{}(\alpha)=(I+\alpha Z)/2$ represents non-unitary operation implemented by measurement along the Z-direction, and it is apparent that the two-qubit unitary channel is decomposed into the one-qubit quantum channels.

Various methods for the quasi-probabilistic decomposition have been proposed~\cite{Mitarai_2021, mitarai2021overhead, Piveteau_2024, Ufrecht_2023, ufrecht2024optimal, Harrow_2025, schmitt2025cutting}. Here, we compute the QCC sampling overhead based on the method in ref.~\cite{Harrow_2025}. This method has several advantages. First, the QCC sampling overhead is optimal. Second, real-time classical communication between partitions is not required. Third, mid-circuit measurement is not required (note that measurement is typically noisier than unitary gate operation). Although ancilla qubits are required, only one ancilla qubit is needed for each partition, independently of the number of decomposed channels.

In this work, we consider that a quantum circuit is cut into two partitions, i.e. bi-partitions~\footnote{While we specifically consider only bi-partitions, our approach in~\ref{sec:gen-cut} may be  easily extended to beyond bi-partitions. For QCC overhead  beyond bi-partitions, see~\cite{Nakamura:2025rlp}.}. As explained in Sec.~\ref{sec:gen-cut}, all operators in $G$ consist of  $R_{ZZ}^{}(\theta)$ and single-qubit gates.
Therefore, it is sufficient to consider the QCC overhead associated with decomposing a group $C$ of  $R_{ZZ}^{}(\theta)$ gates crossing the two partitions, which is~\cite{Harrow_2025}
\begin{align}
  \phi = 2 \prod_{k \in C}\bigl(\bigl|\cos{(\theta_k/2)}\bigr|+\bigl|\sin{(\theta_k/2)}\bigr|\bigr)^2 - 1. \label{eq:phi}
\end{align}
Note that the number of shots $N_{\mathrm{shots}}^{}$ that is required to achieve a accuracy $\epsilon$ in a final output is estimated as $N_{\mathrm{shots}}^{}=O(\phi^2/\epsilon^2)$.

\subsection{GQE under constraints on QCC overhead}\label{sec:gen-cut}

We consider imposing an upper-bound constraint on $\phi$ for all circuits generated by GQE, namely $\phi \le \phi_{\mathrm{max}}$.
As described in Sec.~\ref{sec:gqe}, GQE generates the sequence of operator IDs, $\vec{j} = [0, j_1^{}, j_2^{}, \cdots, j_N^{}]$.
Since each operator in $G$ can be constructed from different number of $R_{ZZ}(\theta)$ with different angle $\theta$, the overhead $\phi$ for the generated circuit corresponding to $\vec{j}$ can be written as
\begin{align}
  \phi = 2 \prod_{j \in \vec{j}} \prod_{k \in C(j)}\bigl(\bigl|\cos{(\theta_k/2)}\bigr|+\bigl|\sin{(\theta_k/2)}\bigr|\bigr)^2 - 1, \label{eq:phi_2}
\end{align}
where $C(j)$ denotes a group of the $R_{ZZ}(\theta)$ gates which consists of the operator $j$ {\it and} subject to the quasi-probabilistic decomposition (i.e. crossing the two partitions).

We introduce an approach that monitors $\phi$ at each step of the sequence generation. An operator can be sampled and appended to the sequence only if the resulting sequence satisfies $\phi \le \phi_{\mathrm{max}}$. To make this approach simpler, we manipulate Eq.~\ref{eq:phi_2} and $\phi \le \phi_{\mathrm{max}}$ as
\begin{align}
  u_j^{} = \sum_{k \in C(j)} \ln{\bigl(\bigl|\cos{(\theta_k/2)}\bigr|+\bigl|\sin{(\theta_k/2)}\bigr|\bigr)}, \label{eq:u_j} \\
  u_{\mathrm{max}} = \frac{1}{2} \ln{\frac{\phi_{\mathrm{max}}+1}{2}}, \\
  \sum_{j \in \vec{j}} u_j^{} \le u_{\mathrm{max}}^{}. \label{eq:const_u}
\end{align}
We also define
\begin{align}
  u_{\mathrm{allowed}}^{}(t) := u_{\mathrm{max}} - \sum_{j \in \vec{j}[:t]} u_j^{},
\end{align}
where $t$ denotes a step index, $t \in \{1, 2, \cdots, N\}$. Note that $\vec{j}[:t]$ does not include $j_t^{}$, i.e. $\vec{j}[:t] = [0, j_1^{}, j_2^{}, \cdots, j_{t-1}^{}]$. Therefore, $u_{\mathrm{allowed}}^{}(t)$ represents the amount of overhead that can still be added at step $t$.
At each step $t$, we disallow operators with overhead above $u_{\mathrm{allowed}}^{}(t)$. Specifically, we assign a zero generation probability to all operators $j \in G$ which satisfy $u_j^{} > u_{\mathrm{allowed}}^{}(t)$. This can be achieved by assigning $-\infty$ to the corresponding logits before applying the softmax function.
This approach is efficient because it does not require  any rejection procedure, such as repeatedly sampling operators and vetoing those that violate the constraint.
Note that $u_j^{}$ can be computed and cached in advance for all operators $j \in G$ once the cutting position is determined, so computing it at every step is unnecessary.

\begin{figure}[t]
\begin{center}
\includegraphics[width=9cm]{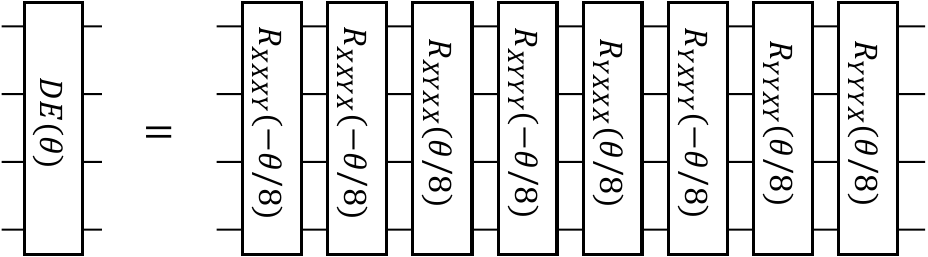}
\end{center}
\caption{A \texttt{qml.DoubleExcitation} operator in PennyLane~\cite{bergholm2022pennylaneautomaticdifferentiationhybrid} is equivalent to eight commuting four-qubit Pauli rotations.}
\label{fig:de}
\end{figure}

\begin{figure}[t]
\begin{center}
\includegraphics[width=4cm]{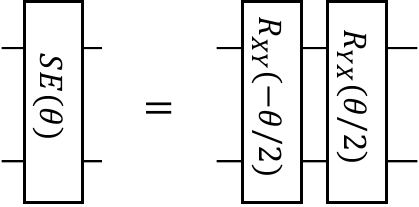}
\end{center}
\caption{A \texttt{qml.SingleExcitation} operator in PennyLane~\cite{bergholm2022pennylaneautomaticdifferentiationhybrid} is equivalent to two commuting two-qubit Pauli rotations.}
\label{fig:se}
\end{figure}

\begin{figure}[t]
\begin{center}
\includegraphics[width=12cm]{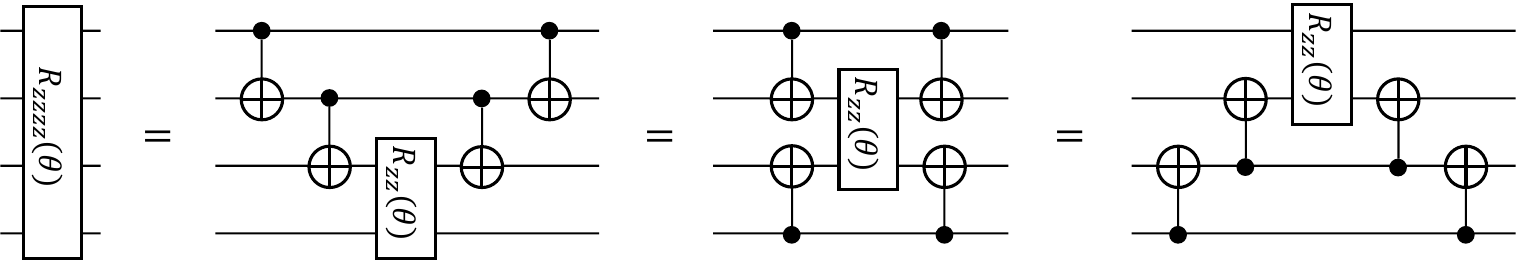}
\end{center}
\caption{A four-qubit Z rotation gate can be written as one two-qubit Z rotation gate sandwiched by ladders of CNOT gates.}
\label{fig:rzzzz}
\end{figure}

As also described in Sec.~\ref{sec:gqe}, we use  \texttt{qml.DoubleExcitation} and \texttt{qml.SingleExcitation} in PennyLane~\cite{bergholm2022pennylaneautomaticdifferentiationhybrid} as operators in $G$, hereafter $DE(\theta)$ and $SE(\theta)$, respectively.
$DE(\theta)$ is a four-qubit operator performing a $SO(2)$ rotation in the two-dimensional subspace $\{\ket{1100}, \ket{0011}\}$~\cite{bergholm2022pennylaneautomaticdifferentiationhybrid},
\begin{subequations}\label{eq:DE}
\begin{align}
  DE(\theta) \ket{1100} = \cos{(\theta/2)} \ket{1100} - \sin{(\theta/2)} \ket{0011}, \\
  DE(\theta) \ket{0011} = \cos{(\theta/2)} \ket{0011} + \sin{(\theta/2)} \ket{1100}.
\end{align}
\end{subequations}
In appendix~\ref{sec:appendix}, we show that $DE(\theta)$ can be given by eight commuting four-qubit Pauli rotations, as shown in Fig.~\ref{fig:de}.
Each of these four-qubit Pauli rotations is equivalent to one four-qubit Z rotation gate, $R_{ZZZZ}^{}(\pm \theta/8)$, up to single-qubit gates. Furthermore, a $R_{ZZZZ}^{}(\theta)$ gate can be written as one $R_{ZZ}^{}(\theta)$ gate sandwiched by ladders of CNOT gates~\cite{Cowtan_2020, ufrecht2024optimal} as shown in Fig.~\ref{fig:rzzzz}.
Therefore, the QCC overhead for decomposing one $DE(\theta)$ operator into two parts is equivalent to that for decomposing a group of eight $R_{ZZ}^{}(\pm \theta/8)$ gates as
\begin{align}
  u_{j=DE(\theta)}^{} = 8 \ln{\bigl(\bigl|\cos{(\theta/16)}\bigr|+\bigl|\sin{(\theta/16)}\bigr|\bigr)}
\end{align}
in terms of $u_j^{}$ defined in Eq.~\ref{eq:u_j}.
For $\theta = \pi/4$, we find $u_{j=DE(\theta=\pi/4)}^{} \simeq 0.37$, which is similar to that for one CNOT gate, $u_{j=CNOT}^{} \simeq 0.35$.
In ref.~\cite{Anselmetti_2021}, it is shown that one $DE(\theta)$ is constructed from CNOT gates and one-qubit gates, where decomposing at least seven CNOT gates is required for decomposing one $DE(\theta)$ into two parts, resulting in $u_{j=DE(\theta)}^{} \simeq 2.43$. This ensures that our approach to decompose $DE(\theta)$ is much more cutting-friendly for relatively small $\theta$ values.

Next, we consider $SE(\theta)$, which is a two-qubit operator performing a $SO(2)$ rotation in the two-dimensional subspace $\{\ket{10}, \ket{01}\}$~\cite{bergholm2022pennylaneautomaticdifferentiationhybrid},
\begin{subequations}\label{eq:SE}
\begin{align}
  SE(\theta) \ket{10} = \cos{(\theta/2)} \ket{10} - \sin{(\theta/2)} \ket{01}, \\
  SE(\theta) \ket{01} = \cos{(\theta/2)} \ket{01} + \sin{(\theta/2)} \ket{10}.
\end{align}
\end{subequations}
In appendix~\ref{sec:appendix}, we show that $SE(\theta)$ can be given by two commuting two-qubit Pauli rotations, as shown in Fig.~\ref{fig:se}.
Each of these two-qubit Pauli rotations is equivalent to one $R_{ZZ}^{}(\pm \theta/2)$ gate up to single-qubit gates.
As a result, the QCC overhead in terms of $u_j^{}$ for decomposing one $SE(\theta)$ operator is equivalent to that for decomposing a group of  two $R_{ZZ}^{}(\theta/2)$ gates as
\begin{align}
  u_{j=SE(\theta)}^{} = 2 \ln{\bigl(\bigl|\cos{(\theta/4)}\bigr|+\bigl|\sin{(\theta/4)}\bigr|\bigr)}.
\end{align}
For $\theta = \pi/4$, we find $u_{j=SE(\theta=\pi/4)}^{} \simeq 0.32$.
In ref.~\cite{Anselmetti_2021}, it is shown that one $SE(\theta)$ is constructed from CNOT gates and one-qubit gates, where decomposing two CNOT gates is required for decomposing one $SE(\theta)$, resulting in $u_{j=SE(\theta)}^{} \simeq 0.69$. This also ensures that our approach to decompose $SE(\theta)$ is much more cutting-friendly for relatively small $\theta$ values.

\begin{figure}[t]
\begin{center}
\includegraphics[width=8cm]{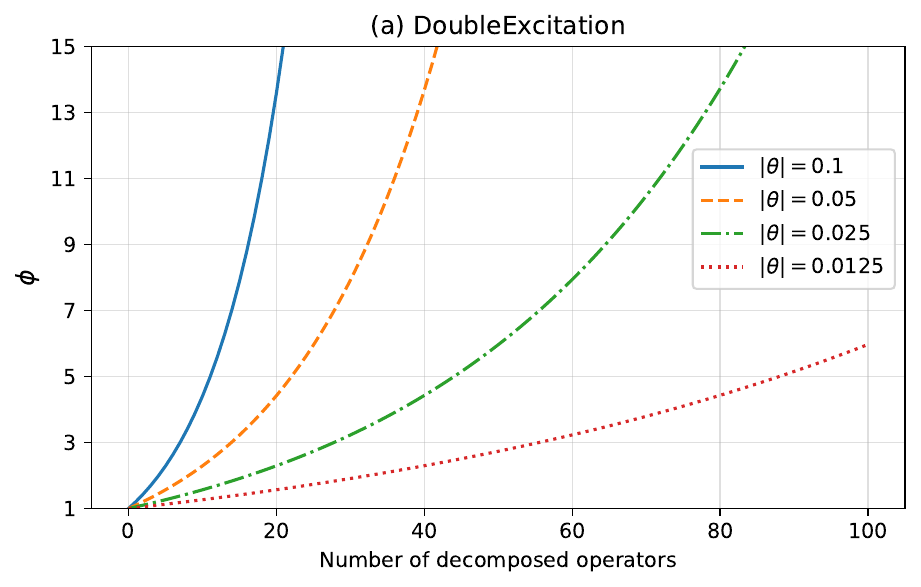}
\includegraphics[width=8cm]{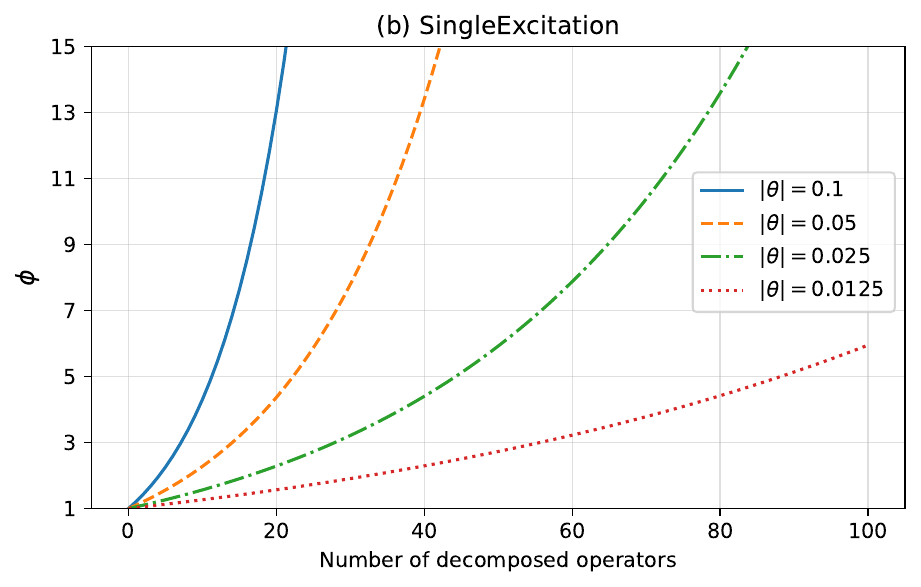}
\end{center}
\caption{QCC overhead $\phi$ defined in Eq.~\ref{eq:phi_2} as a function of the number of decomposed $DE(\theta)$ (left panel) and $SE(\theta)$ (right panel) operators for four different values of the angle $|\theta|$.}
\label{fig:overhead_values}
\end{figure}

In Fig.~\ref{fig:overhead_values} we show the QCC overhead $\phi$ defined in Eq.~\ref{eq:phi_2} as a function of the number of decomposed $DE(\theta)$ (left panel) and $SE(\theta)$ (right panel) operators for four different values of the angle $|\theta|$.
It is observed that the difference in QCC overhead between $DE(\theta)$ and $SE(\theta)$  is small, despite the naive expectation that $DE(\theta)$ has a much larger QCC overhead than $SE(\theta)$.

\begin{figure}[t]
\begin{center}
\includegraphics[width=10cm]{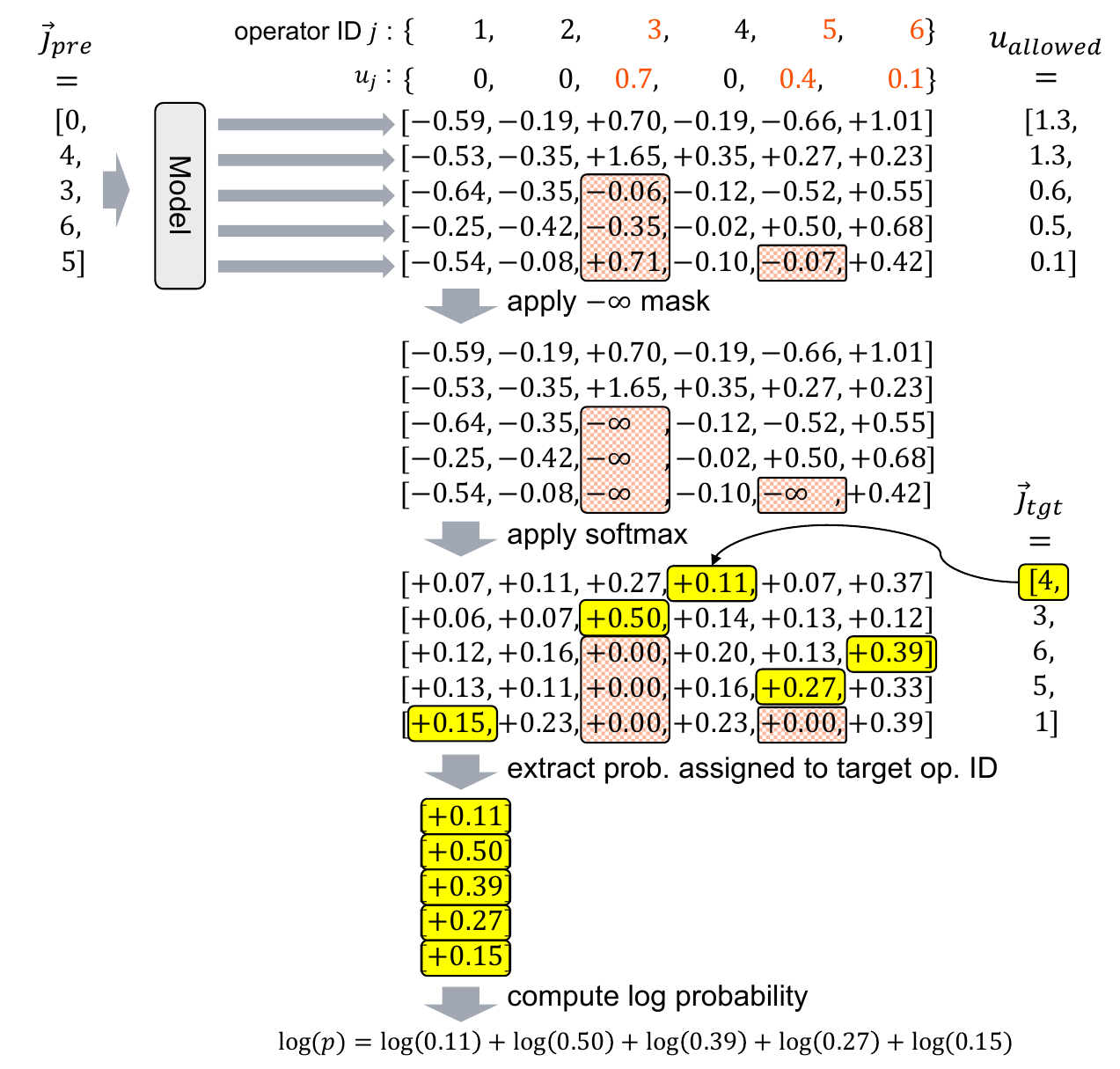}
\end{center}
\caption{Loss-masking procedure to compute $\log{p_{\theta}^{}(\vec{j})}$ for a given operator-ID sequence $\vec{j}=[0, 4, 3, 6, 5, 1]$. The operator pool size is set to $L=6$. It is assumed that $u_{\mathrm{max}}=1.3$ and operator IDs $\{3, 5, 6\}$ are subject to decompositions with incremental overheads $u_{j=3}^{}=0.7$, $u_{j=5}^{}=0.4$, and $u_{j=6}^{}=0.1$, respectively.
The computation mirrors the constrained generation procedure that we disallow the operator ID 3 after step $t=2$ ($t \ge 3$) and  the operator ID 5 after step $t=4$ ($ t \ge 5$), because appending these operators violate the constraint (see Eq.~\ref{eq:const_u}).
Refer to the main text for further details.}
\label{fig:loss}
\end{figure}

When, as in the training phase of GQE, the current model weights determine the next data sample (circuit) and those samples are then used to update the model weights, the process of updating the model weights (more precisely computing the loss function) needs to take into account the same constraint $\phi \le \phi_{\mathrm{max}}$ as the generation process.
Specifically, our loss function defined in Sec.~\ref{sec:pdpo} consists of $\log{p_{\theta}^{}(\vec{j})}$ for a given $\vec{j}$  which satisfies the constraint $\phi \le \phi_{\mathrm{max}}$, and  the computation of such $\log{p_{\theta}^{}(\vec{j})}$ requires the following procedure, which we call {\it loss-masking}.
Assume that $\vec{j} = [0, j_1^{}, j_2^{}, \cdots, j_N^{}]$ is given. First, we construct the target sequence $\vec{j}_{\mathrm{tgt}} = [j_1^{}, j_2^{}, \cdots, j_{N}^{}]$, and the prefix sequence $\vec{j}_{\mathrm{pre}} = [0, j_1^{}, j_2^{}, \cdots, j_{N-1}^{}]$ which is used to predict $\vec{j}_{\mathrm{tgt}}$.
By inputting $\vec{j}_{\mathrm{pre}}$ into the model, we obtain logits corresponding to the probabilities for $\vec{j}_{\mathrm{tgt}}$.
Then, as in the generation process, we compute the remaining allowance $u_{\mathrm{allowed}}^{}(t)$ at each step ($t \in \{1, 2, \cdots, N\}$) from  $\vec{j}$, and assign $-\infty$ to the logits of operators whose incremental overhead (i.e. $u_j^{}$) exceeds this allowance.
After applying the softmax function, we extract the probabilities assigned to $\vec{j}_{\mathrm{tgt}}$.
By summing log of these probabilities, we obtain the log probability of the full sequence $\vec{j}$ as
\begin{align}
  \log{p_{\theta}^{}\bigl(\vec{j}\bigr)} = \sum_{t \in  \{1, 2, \cdots, N\}} \log{p_{\theta}^{}\bigl(j_{t} | \vec{j}_{\mathrm{pre}}[:t]\bigr)}.
\end{align}
In Fig.~\ref{fig:loss}, we illustrate the loss-masking procedure to compute $\log{p_{\theta}^{}(\vec{j})}$ for $\vec{j}=[0, 4, 3, 6, 5, 1]$ in a simplified case of $L=6$ and $N=5$. The loss-masking procedure ensures that the computation of $\log{p_{\theta}^{}(\vec{j})}$ is consistent with the constrained generation procedure for $\vec{j}$ under $\phi \le \phi_{\mathrm{max}}$.

\section{New loss function and hybrid approach}\label{sec:pdpo_hbd}

In this section, we introduce two new methods to improve GQE.
In \ref{sec:pdpo}, we first  briefly review direct preference optimization (DPO), identify the limitation of DPO when used as the loss function in GQE, and introduce Persistent-DPO (P-DPO) as a solution to this limitation.
In \ref{sec:hybrid} we explain the hybrid approach that combines offline and online learning during the training phase.
Note that these two methods are applicable regardless of whether the constrained circuit generation described in Sec.~\ref{sec:gqe_qcc} is employed.

\subsection{Persistent DPO objective}\label{sec:pdpo}

DPO~\cite{rafailov2024directpreferenceoptimizationlanguage} has been proposed as a method to align a pre-trained LLM with human preferences using simple supervised learning. Its loss function  is given by
\begin{align}
  {\cal L}_{\mathrm{DPO}} & = - \mathbb{E}_{(\vec{j}_w^{}, \vec{j}_l^{}) \sim {\cal B}} \bigl[ \log\sigma(z)   \bigr], \\
  z & = \beta \log{\frac{p_{\theta}^{}(\vec{j}_w^{})}{p_{\mathrm{ref}}^{}(\vec{j}_w^{})}} - \beta \log{\frac{p_{\theta}^{}(\vec{j}_l^{})}{p_{\mathrm{ref}}^{}(\vec{j}_l^{})}},\label{eq:z}
\end{align}
where $p_{\mathrm{ref}}^{}$ denotes a reference model that is not subject to update, $\sigma$ is the Sigmoid function, and $\beta(>0)$ is a hyperparameter.
Note that $z$ is the difference between the two functions as
\begin{align}
  z = r_{\theta}^{}(\vec{j}_w^{}) - r_{\theta}^{}(\vec{j}_l^{}),
\end{align}
where $r$ denotes the DPO's reward formulation (the partition function is excluded)
\begin{align}
  r_{\theta}^{}(\vec{j}) = \beta \log{\frac{p_{\theta}^{}(\vec{j})}{p_{\mathrm{ref}}^{}(\vec{j})}}.
\end{align}
The loss function is computed for a pair of two samples denoted by $\vec{j}_w^{}$ and $\vec{j}_l^{}$. Here $\vec{j}_w^{}$ denotes the preferable sample over $\vec{j}_l^{}$ in terms of the expectation value of a target Hamiltonian~\cite{minami2025generative}, namely $E(\vec{j}_w^{}) < E(\vec{j}_l^{})$. The dataset ${\cal B}$ denotes a batch ${\cal B} = \bigl\{(\vec{j}_w^{(i)}, \vec{j}_l^{(i)})\bigr\}_{i=1}^{M_B^{}}$, where $M_B^{}$ is the batch size. The gradient of ${\cal L}_{\mathrm{DPO}}$ with respect to the model weights $\theta$ can be written as~\cite{rafailov2024directpreferenceoptimizationlanguage},
\begin{align}
  \nabla_{\theta}^{} {\cal L}_{\mathrm{DPO}} = - \beta \mathbb{E}_{(\vec{j}_w^{}, \vec{j}_l^{}) \sim {\cal B}} \biggl[ \sigma(-z)  \Bigl[ \nabla_{\theta}^{} \log{p_{\theta}^{}(\vec{j}_w^{})} - \nabla_{\theta}^{} \log{p_{\theta}^{}(\vec{j}_l^{})}  \Bigr] \biggr].\label{eq:gradient_dpo}
\end{align}
When the model assigns a low probability to $\vec{j}_w^{}$ and a high probability to $\vec{j}_l^{}$ (which is a wrong estimate), $z$ takes a large value in the negative direction, as apparent from Eq.~(\ref{eq:z}).
In such a case, the weighting coefficient $\sigma(-z)$ in the gradient can approach its maximum value $\sigma(-z) \simeq 1$, and therefore the model weights are largely updated to strongly correct the wrong estimate.
As the opposite case, when the model assigns a high probability to $\vec{j}_w^{}$ and a low probability to $\vec{j}_l^{}$ (which is a correct estimate), $z$ takes a large value in the positive direction. In such a case, the weighting coefficient $\sigma(-z)$ can vanish $\sigma(-z) \simeq 0$, and therefore the model weights are little updated.
This makes sense when using DPO for the alignment of a pre-trained LLM, because the model being updated is already pre-trained and it may be preferable not to update it significantly when it has already correctly estimated the preference for a given pair.
It has been actually reported in ref.~\cite{rafailov2024directpreferenceoptimizationlanguage} that the models collapsed without $\sigma(-z)$.

However, when DPO is used for GQE, the weighting coefficient $\sigma(-z)$ may limit learning. This is  because the model weights should be persistently updated even for sample pairs whose preferences are already correctly estimated by the model, so that it can generate even more preferable circuits.
This should be particularly true in the pre-training phase and in the training phase from scratch. Even in the training phase starting from the pre-trained model, this is considered to be true, because its purpose  is not the alignment.

In order to prevent the gradient from vanishing for a positively large $z$, we introduce P-DPO, whose loss function is defined by
\begin{align}
    {\cal L}_{\mathrm{P-DPO}} = - \mathbb{E}_{(\vec{j}_w^{}, \vec{j}_l^{}) \sim {\cal B}} \bigl[ \alpha z + (1-\alpha)\log\sigma(z) \bigr], \\
    0 \le \alpha \le 1,
\end{align}
where $z$ is the same as Eq.~(\ref{eq:z}).
It is straightforward to derive the gradient of ${\cal L}_{\mathrm{P-DPO}}$ with respect to the model weights $\theta$,
\begin{align}
  \nabla_{\theta}^{} {\cal L}_{\mathrm{P-DPO}} & = - \beta \mathbb{E}_{(\vec{j}_w^{}, \vec{j}_l^{}) \sim {\cal B}} \biggl[ w(z) \Bigl[ \nabla_{\theta}^{} \log{p_{\theta}^{}(\vec{j}_w^{})} - \nabla_{\theta}^{} \log{p_{\theta}^{}(\vec{j}_l^{})}  \Bigr] \biggr], \label{eq:gradient_pdpo} \\
  w(z) & = \alpha + (1-\alpha)\sigma(-z).
\end{align}
The new weighting coefficient $w(z)$ has a value in the range $\alpha \le w(z) \le 1$. It is lower bounded by $\alpha$ even for $z \to +\infty$. P-DPO reduces to DPO as $\alpha \to 0$.
When $\alpha = 1$, the weighting coefficient becomes independent of $z$.
By adopting this loss function with an appropriate value for hyperparameter $\alpha$, it becomes possible to persistently apply a minimal learning pressure even to sample pairs whose preferences are already correctly estimated by the generative model.

We note that another approach to suppress the effect of  $\sigma(-z)$ is to assign a very small value to $\beta$ in Eq.~(\ref{eq:z}), so that $\sigma(-z) \simeq 0.5$ regardless of the value of $z$ apart from $\beta$.
However, this approach also suppresses the desirable property that the model parameters are relatively strongly updated for wrongly estimated sample pairs.
Even another approach is to assign an appropriate value to $\beta$ so that the maximum value of $z$ is upper-bounded by $\sim 1$. However, finding the appropriate value in advance is a difficult task, because it depends on experimental setup as follows.
Each log-probability in Eq.~(\ref{eq:z}) for example $\log{p_{\theta}^{}(\vec{j}_w^{})}$ is the sum of the log-probability for each operator ID,
\begin{align}
  \log{p_{\theta}^{}(\vec{j}_w^{})} = \sum_{i=1}^N \log{p_{\theta}^{}(j_i^{}\ |\ j_{<i}^{} )}.\label{eq:logprob}
\end{align}
This indicates that the value of $z$ apart from $\beta$ tends to increase as the length $N$ of the sequence and/or the size $L$ of the operator pool $G$ are set large~\cite{meng2024simposimplepreferenceoptimization}.

\subsection{Hybrid approach combining offline and online learning}\label{sec:hybrid}

Our approach is based on the idea of experience replay~\cite{lin1993reinforcement, mnih2015human} utilized in the field of reinforcement learning.
Our specific algorithm is as follows.
At each iteration step $t$, we store the online dataset $d_t^{} = \bigl\{(\vec{j}_{}^{(t, i)}, E(\vec{j}_{}^{(t, i)})) \bigr\}_{i=1}^{M}$ that the generative model under learning produced, in a data set $D_t^{} = \{d_1^{}, \dots, d_t^{} \}$. The dataset $D_t^{}$ has the fixed capacity $C$. If the size of $D_t^{}$ exceeds $C$, it reduces to $C$ in such a way that samples with the highest energy are removed first.
Our algorithm has a parameter $R$ that specifies the number of samples reused at each iteration step.
Specifically, at each iteration step $t$, we randomly select $R$ samples from $D_t^{}$. Those samples are added to the online dataset $d_t^{}$, resulting in the $M+R$ samples used to update the model.
Our algorithm also introduces a parameter $S$ that specifies the iteration step at which the hybrid approach starts. Even before the step $S$ the dataset $D_t^{}$ is constructed, and the reuse from $D_t^{}$ occurs after the step $S$.
Therefore, our algorithm has three hyperparameters in total, capacity $C$, reuse size $R$ and starting point $S$.

The difference between our algorithm and experience replay in ref.~\cite{mnih2015human} is that, the online dataset $d_t^{}$ is always included in the samples for a model update at step $t$ in our method, whereas in experience replay this occurs randomly.
In addition, when the size of $D_t^{}$ exceeds $C$, our method deletes samples from those with the highest energy, while experience replay removes the oldest entries.

\section{Numerical studies}\label{sec:numerical}

\subsection{Setup}\label{sec:setup}

\begin{table}[t]
\begin{tabular}{p{25em}p{10em}}
\toprule
  Hyperparameter & Value \\
\midrule
 Dimension of embedding & 768 \\
 Dimension of feed-forward layer & 3072 \\
 Number of attention heads & 12 \\
 Number of layers & 12 \\
 Bias & False \\
 Activation function & GELU \\
 Optimizer & AdamW \\
 Learning rate & $8 \times 10^{-5}$ \\
 Weight decay & $0.01$ \\
 Dropout rate & 0 \\
 Length $N$ of sequence & 50 \\
  Number of circuits generated per step & 10 \\
 Random seeds & 42, 123, 777, 2024, 9999 \\
\bottomrule
\end{tabular}
\caption{List of hyperparameters not described in the main text.}
\label{tbl:hyper}
\end{table}

\begin{figure}[t]
\begin{center}
\includegraphics[width=5cm]{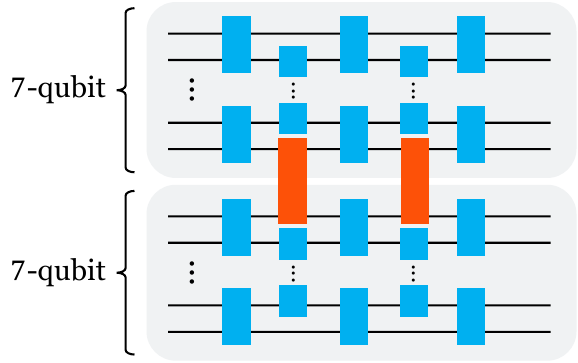}
\end{center}
\caption{An illustration of a quantum circuit generated by GQE under constraints on QCC overhead. The circuit consists of two partitions, and distinguishes between operators within the partition (blue) and those crossing the partitions (red). The latter operators contribute to QCC overhead, and therefore are constrained.}
\label{fig:circuit}
\end{figure}

We implement GQE with a transformer decoder of the GPT-2 architecture~\cite{vaswani2017attention, radford2019language} following ref.~\cite{nakaji2024generative}.
The transformer decoder is implemented using PyTorch~\cite{paszke2019pytorchimperativestylehighperformance} with help of ref.~\cite{build-llms-from-scratch-book}.
As a target Hamiltonian, we use the Hamiltonian of the $\mathrm{BeH_2^{}}$ molecule in the sto-3g basis with a bond length of $2.1\text{\AA}$. Specifically, we use the dataset from ref.~\cite{Utkarsh2023Chemistry}.
Our operator pool $G$ consists of $DE(\theta)$ and $SE(\theta)$ operators derived from the target molecule, where $\theta$ of these operators take values from $\{\pm 2^k_{}/160 \}_{k=1}^4$. As a result, the size of $G$ is $L=1633$ including the identity operator.
As the initial state on which the operators specified by $\vec{j}$ are applied, we use the Hartree-Fock state.
We consider that 14-qubit quantum circuits are cut into two equal size partitions as illustrated in Fig.~\ref{fig:circuit}, where one partition has the even-numbered 7-qubit (i.e. $0, 2, \dots, 12$) and the other partition has the odd-numbered 7-qubit (i.e. $1, 3, \dots, 13$). As a result, among the 1633 operators in the gate pool, the 481 operators are within the partitions and the remaining 1152 operators are crossing the partitions.

When computing the loss function of DPO or P-DPO, we have to make a pair $(\vec{j}_w^{}, \vec{j}_l^{})$ of samples. As the pairing strategy, we adopt the best-vs-others proposed in ref.~\cite{minami2025generative}.
We focus on the training phase from scratch, i.e, without the pre-training phase. The model $p_{\theta}^{}$ is randomly initialized.
There is some arbitrariness in the choice of the reference model $p_{\mathrm{ref}}^{}$ in Eq.~(\ref{eq:z}), particularly when the pre-training phase is absent.
We fix $p_{\mathrm{ref}}^{}$ to the randomly initialized model for $p_{\theta}^{}$ at the beginning of training and keep it constant throughout the training phase.
We note that, when the pre-training phase is conducted before the training phase, it is natural to use the pre-trained model as the reference model.
We use $\beta=0.1$ used in refs.~\cite{rafailov2024directpreferenceoptimizationlanguage,minami2025generative}.

The temperature $T$ for temperature scaling is initially set high and gradually decreased as the iterations proceed. At iteration step $i$, we use $T=T_{\mathrm{initial}}^{}-(T_{\mathrm{initial}}^{}-T_{\mathrm{final}}^{})\times i/(N_{\mathrm{steps}^{}}-1)$, where $T_{\mathrm{initial}}^{}$ denotes the temperature at the beginning ($i=0$), $T_{\mathrm{final}}^{}$ denotes the temperature at the end ($i=N_{\mathrm{steps}^{}}-1$), and $N_{\mathrm{steps}^{}}$ denotes the total number of iterations. We set $T_{\mathrm{initial}}^{}=1.5$, $T_{\mathrm{final}}^{}=0.7$ and $N_{\mathrm{steps}^{}}=3000$.

All executions of quantum circuits and calculations of the expectation values of the Hamiltonian are classically performed using PennyLane~\cite{bergholm2022pennylaneautomaticdifferentiationhybrid} \texttt{version 0.42.1} with lightning plugins~\cite{asadi2024}.
As the purpose of this study is to first evaluate the effectiveness of our proposed methods under ideal conditions, we consider neither shot noise nor decoherence.
Other relevant parameters including hyperparameters in the transformer decoder  are summarized in Table~\ref{tbl:hyper}~\footnote{The codes are available from the authors upon request.}.

\subsection{Result}\label{sec:result}

\begin{figure}[t]
\begin{center}
\includegraphics[width=8cm]{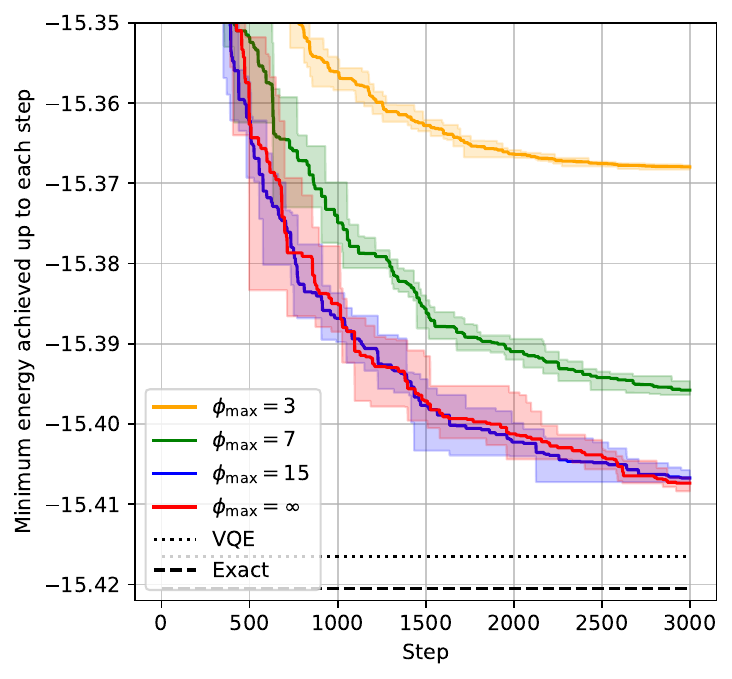}
\end{center}
\caption{The minimum energy values in unit of Hartree achieved up to each iteration step are plotted as a function of the iteration step.
The upper-bound constraint $\phi \le \phi_{\mathrm{max}}$ is imposed on all circuits generated by GQE, and the results with four different values of $\phi_{\mathrm{max}}$ are shown.
The mean value over the five runs with five different random seeds is shown by a solid line, and the region spanning the maximum and minimum values of those five runs is shaded. The black dashed line shows the ground state energy from the exact calculation, and the black dotted line shows the VQE energy from the PennyLane dataset~\cite{Utkarsh2023Chemistry}.}
\label{fig:acc_min_energy_phi_dependence}
\end{figure}

\begin{figure}[t]
\begin{center}
\includegraphics[width=8cm]{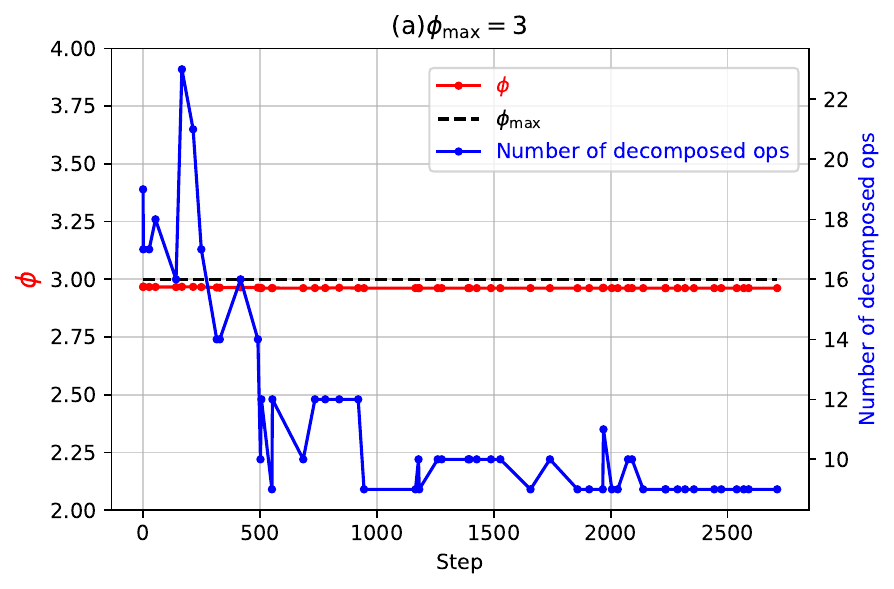}
\includegraphics[width=8cm]{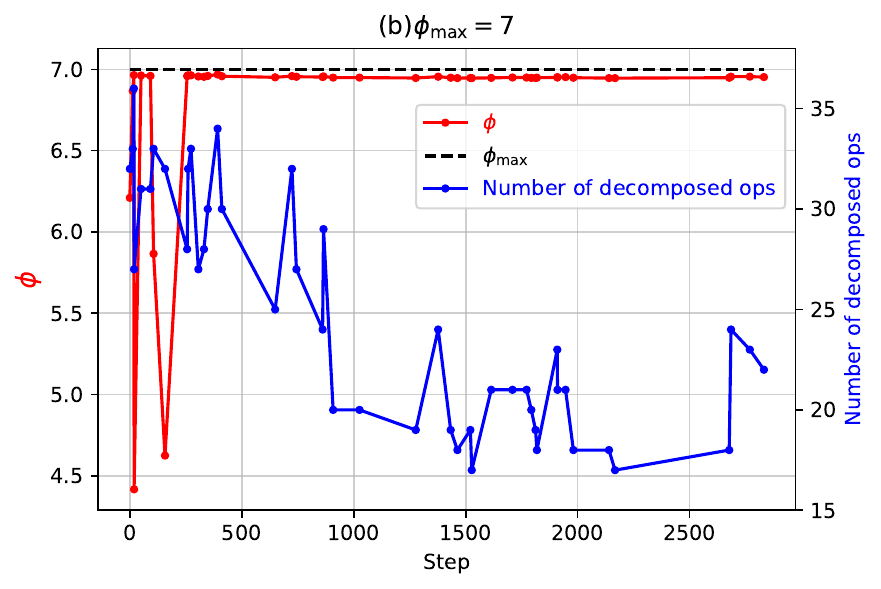}
\includegraphics[width=8cm]{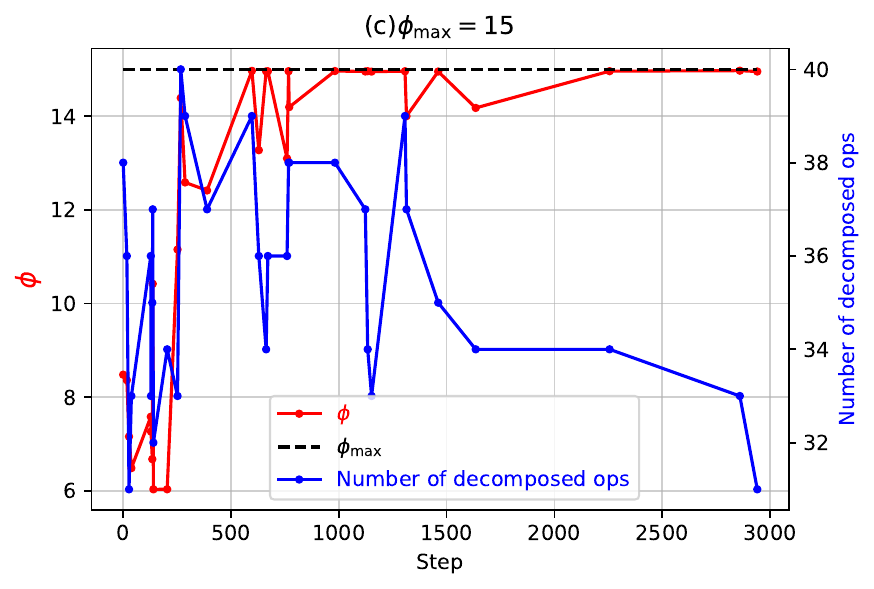}
\includegraphics[width=8cm]{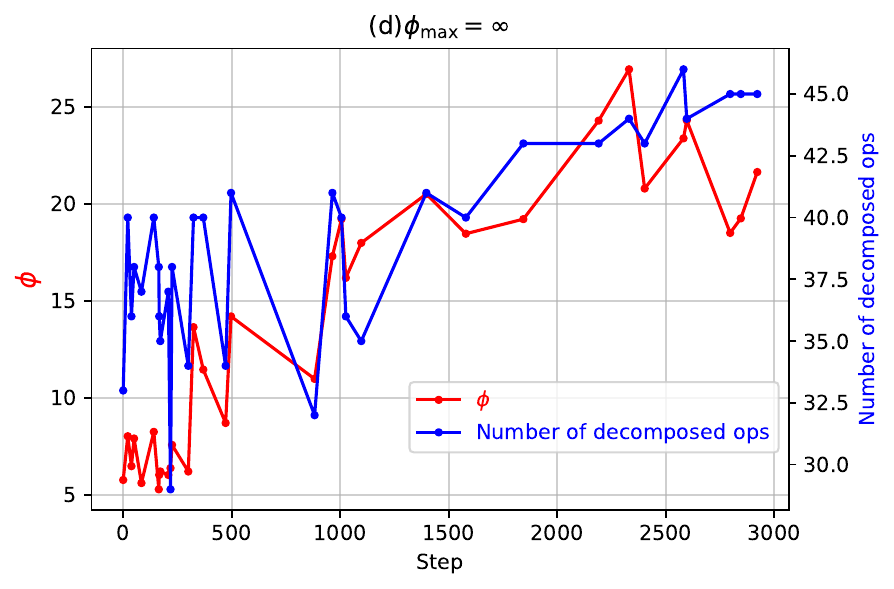}
\end{center}
\caption{$\phi$ on the left vertical axis and the number of decomposed $DE(\theta)$ and $SE(\theta)$ operators on the right vertical axis as functions of the iteration step.
The results of $\phi_{\mathrm{max}}=3$, $\phi_{\mathrm{max}}=7$, $\phi_{\mathrm{max}}=15$ and $\phi_{\mathrm{max}}=\infty$ (i.e. no constraint) are shown in the panel (a), (b), (c) and (d), respectively.
We select only the seed result that achieves the best minimum energy among those five different seed results shown  in Fig.~\ref{fig:acc_min_energy_phi_dependence}.
Both  $\phi$ and  the number of decomposed operators are plotted only at iteration steps where the minimum energy achieved up to that step is updated. The black dashed line shows $\phi_{\mathrm{max}}$.}
\label{fig:overhead_phimax}
\end{figure}

At first, we study the impact of the constraint $\phi \le \phi_{\mathrm{max}}$.
In Fig.~\ref{fig:acc_min_energy_phi_dependence}, we plot the minimum energy values in unit of Hartree achieved up to each iteration step as a function of the iteration step. The results for four different values of $\phi_{\mathrm{max}}$ are shown, $\phi_{\mathrm{max}}=3$ (yellow curve), $\phi_{\mathrm{max}}=7$ (green curve), $\phi_{\mathrm{max}}=15$ (blue curve) and $\phi_{\mathrm{max}}=\infty$, i.e. no constraint (red curve). Note that $\phi=3$,  $\phi=7$, and $\phi=15$ correspond to the overheads for decomposing one, two, and three CNOT gates, respectively.
To study a random seed dependence, each experiment is conducted five times by initializing the generative model and the simulator with five different random seeds. The mean value over the five runs with five different random seeds is shown by a solid line, and the region spanning the maximum and minimum values of those five runs is shaded. The black dashed line shows the ground state energy from the exact calculation, and the black dotted line shows the VQE energy from the PennyLane dataset~\cite{Utkarsh2023Chemistry}. Note that DPO is used as the loss function and the hybrid approach in Section~\ref{sec:hybrid} is not used for these results.

In Fig.~\ref{fig:overhead_phimax}, we plot $\phi$ on the left vertical axis and the number of decomposed $DE(\theta)$ and $SE(\theta)$ operators on the right vertical axis as functions of the iteration step. Here, we select only the seed result that achieves the best minimum energy among those five different seed results shown  in Fig.~\ref{fig:acc_min_energy_phi_dependence}.
Both  $\phi$ and  the number of decomposed operators are plotted only at iteration steps where the minimum energy achieved up to that step is updated. The black dashed line shows $\phi_{\mathrm{max}}$.

It is observed in Fig.~\ref{fig:overhead_phimax} that the constraint $\phi \le \phi_{\mathrm{max}}$ is always satisfied as anticipated. When the constraint $\phi \le \phi_{\mathrm{max}}$ is absent, as in panel (d) of Fig.~\ref{fig:overhead_phimax}, $\phi$ increases beyond $20$. Nevertheless, the difference in the minimum energy values between $\phi_{\mathrm{max}}=15$ and $\phi_{\mathrm{max}}=\infty$ is quite small, as observed in Fig.~\ref{fig:acc_min_energy_phi_dependence}.
This observation indicates that GQE explores low-energy solutions that satisfy the constraint $\phi \le \phi_{\mathrm{max}}$, as expected.

\begin{figure}[t]
\begin{center}
\includegraphics[width=8cm]{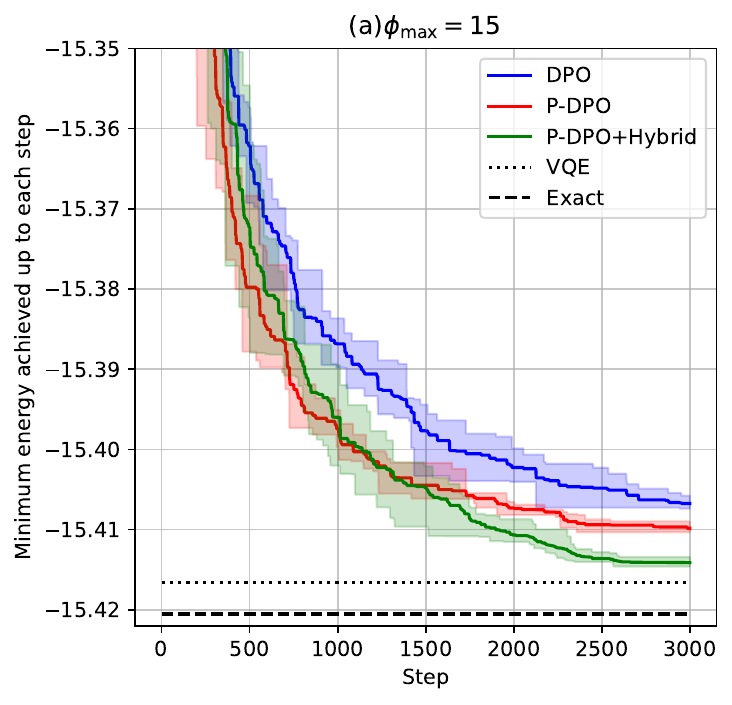}
\includegraphics[width=8cm]{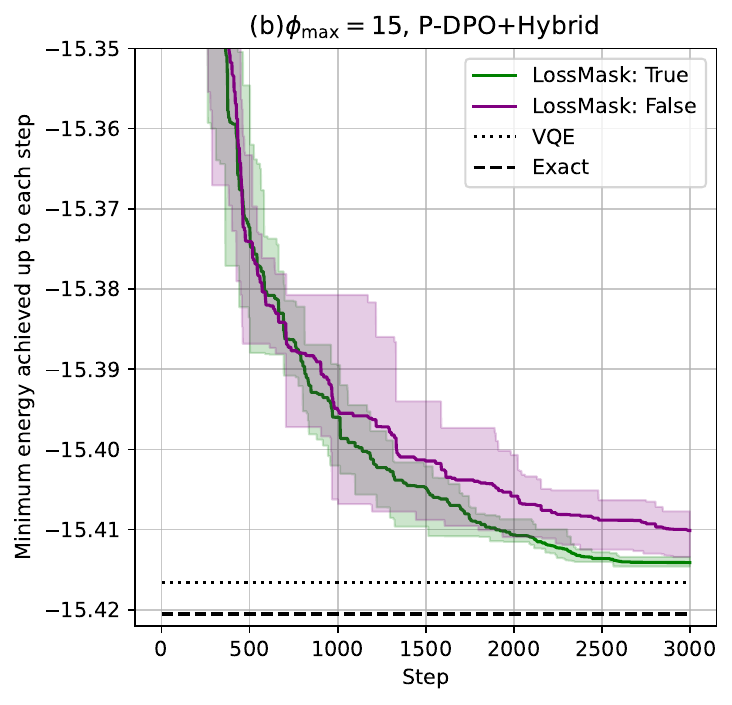}
\end{center}
\caption{(a) DPO, P-DPO ($\alpha=0.3$) and P-DPO ($\alpha=0.3$) combined with the hybrid approach  ($(C, R, S)=(100, 2, 50)$) are compared under the constraint $\phi \le \phi_{\mathrm{max}}=15$ in the same manner as Fig.~\ref{fig:acc_min_energy_phi_dependence}. (b) The result with the loss-masking (labeled as True) and that without the loss-masking (labeled as False) are compared in the same manner as Fig.~\ref{fig:acc_min_energy_phi_dependence}, where   the constraint $\phi \le \phi_{\mathrm{max}}=15$ and P-DPO ($\alpha=0.3$) combined with the hybrid approach ($(C, R, S)=(100, 2, 50)$) are commonly used.}
\label{fig:pdpo_hybrid}
\end{figure}

Next, we evaluate P-DPO and the hybrid approach.  In the panel (a) of Fig.~\ref{fig:pdpo_hybrid}, DPO, P-DPO, and P-DPO combined with the hybrid approach are compared under the constraint $\phi \le \phi_{\mathrm{max}}=15$ in the same manner as Fig.~\ref{fig:acc_min_energy_phi_dependence}. The parameter $\alpha$ in P-DPO is set to $\alpha=0.3$. The parameters in the hybrid approach are set to $(C, R, S)=(100, 2, 50)$. It is observed that P-DPO achieves the lower energy values than DPO, and that the further improvement is obtained when combined with the hybrid approach.

Finally, we evaluate the effectiveness of the loss-masking introduced in Section~\ref{sec:gen-cut}. In the panel (b) of Fig.~\ref{fig:pdpo_hybrid}, the result with the loss-masking (labeled as True) and that without the loss-masking (labeled as False) are compared in the same manner as Fig.~\ref{fig:acc_min_energy_phi_dependence}, where   the constraint $\phi \le \phi_{\mathrm{max}}=15$ and P-DPO ($\alpha=0.3$) combined with the hybrid approach ($(C, R, S)=(100, 2, 50)$) are commonly used.
It is observed that the result with the loss-masking exhibits better energy convergence and greater stability across random seeds, thus validating its effectiveness.

\section{Conclusion}\label{sec:conclusion}

We have extended GQE such that the classical generative model produces only quantum circuits whose QCC overhead is upper-bounded while retaining the original purpose of GQE, namely generating circuits with desired properties such as approximating molecular ground states.
We have implemented GQE with a transformer decoder and numerically validated our approach through the simulated ground state search experiments on the $\mathrm{BeH_2^{}}$ molecule.

Specifically, we have introduced an approach that monitors the QCC overhead $\phi$ at each step of sequential  circuit generation such that an operator is allowed to be sampled and appended to the circuit only if the resulting circuit satisfies the constraint $\phi \le \phi_{\mathrm{max}}$.
This approach has been validated by the results shown in Figs.~\ref{fig:acc_min_energy_phi_dependence} and~\ref{fig:overhead_phimax}. In particular, we have observed that the minimum energy achieved by GQE under the constraint $\phi \le \phi_{\mathrm{max}}=15$ is nearly identical to that without the constraint, while the former case yields a smaller QCC overhead due to the imposed constraint; see the blue and red curves in Fig.~\ref{fig:acc_min_energy_phi_dependence} and panels (c) and (d) of Fig.~\ref{fig:overhead_phimax}.
Also, the method to achieve better energy convergence and greater stability across random seeds has been introduced (named loss-masking) and validated by the results shown in the panel (b) of Fig.~\ref{fig:pdpo_hybrid}.

We have also introduced the two methods to improve GQE, Persistent-DPO and the hybrid approach that combines offline and online learning.
We have observed that these methods contribute to faster convergence to low energies; see the panel (a) of Fig.~\ref{fig:pdpo_hybrid}.

The numerical results in this work are limited to approaching the solution achieved by VQE (retrieved from the PennyLane dataset).  Our operator pool $G$ consists of the same VQE ans\"atze~\cite{Anselmetti_2021} that is employed to produce this VQE solution. Therefore, as long as this operator pool is employed, this energy value might be the limit even with extensive hyperparameter tuning. Reaching the exact solution may require the development of ans\"atze specially designed for GQE, which we leave as future work.

\section*{Acknowledgments}
The views expressed in this research are those of the authors and do not necessarily reflect the official policy or position of PwC Consulting LLC.
We wish to thank Mitsuhiro Matsumoto, Tsuyoshi Kitano, Takahiko Satoh, Hana Ebi, Shin Nishio and Takaharu Yoshida for stimulating discussion and support. This work is supported by New Energy and Industrial Technology Development Organization (NEDO).

\bibliography{main}
\bibliographystyle{unsrt}

\appendix
\section{Double- and single-excitation operators in cutting-friendly forms}\label{sec:appendix}

In this appendix, we show that $DE(\theta)$ and $SE(\theta)$ can be given by eight commuting four-qubit Pauli rotations and two commuting two-qubit Pauli rotations, respectively, as shown in Figs.~\ref{fig:de} and~\ref{fig:se}.

We introduce the ladder operators $\sigma^{\pm}_{} := (X\mp iY)/2$, which satisfy
\begin{align}
  \sigma^+_{} \ket{0} = \ket{1},\ \ \sigma^+_{} \ket{1} = 0,\ \ \sigma^-_{} \ket{0} = 0,\ \ \sigma^-_{} \ket{1} = \ket{0}.
\end{align}
With the ladder operators, we achieve
\begin{align}
  \sigma_0^- \sigma_1^- \sigma_2^+ \sigma_3^+ \ket{1100}  = \ket{0011},\ \
  \sigma_0^- \sigma_1^- \sigma_2^+ \sigma_3^+ \ket{\mathrm{other\ than}\ 1100}  = 0, \\
  \sigma_0^+ \sigma_1^+ \sigma_2^- \sigma_3^- \ket{0011}  = \ket{1100},\ \
  \sigma_0^+ \sigma_1^+ \sigma_2^- \sigma_3^- \ket{\mathrm{other\ than}\ 0011}  = 0.
\end{align}
Therefore, if we define $H := i (\sigma_0^+ \sigma_1^+ \sigma_2^- \sigma_3^- - \sigma_0^- \sigma_1^- \sigma_2^+ \sigma_3^+)$, which is hermitian, it is straightforward to confirm that the following satisfies Eq.~\ref{eq:DE},
\begin{align}
  DE(\theta) = I \cos{(\theta/2)} - i H \sin{(\theta/2)} = e^{-i\frac{\theta}{2}H}.
\end{align}
Now we decompose $H$ in terms of $X$ and $Y$ as
\begin{align}
  H = \frac{1}{8}\bigl(-X_0^{}X_1^{}X_2^{}Y_3^{}-X_0^{}X_1^{}Y_2^{}X_3^{}+X_0^{}Y_1^{}X_2^{}X_3^{}-X_0^{}Y_1^{}Y_2^{}Y_3^{}\nonumber \\
  +Y_0^{}X_1^{}X_2^{}X_3^{}-Y_0^{}X_1^{}Y_2^{}Y_3^{}+Y_0^{}Y_1^{}X_2^{}Y_3^{}+Y_0^{}Y_1^{}Y_2^{}X_3^{}\bigr),
\end{align}
where all these eight terms commute with each other. Therefore, $DE(\theta)$ can be given by eight commuting four-qubit Pauli rotations.

Next, we consider $SE(\theta)$.
As in $DE(\theta)$, we define a hermitian operator $K := i (\sigma_0^+ \sigma_1^- - \sigma_0^- \sigma_1^+)$, where each term satisfies
\begin{align}
  \sigma_0^- \sigma_1^+ \ket{10}  = \ket{01},\ \
  \sigma_0^- \sigma_1^+ \ket{\mathrm{other\ than}\ 10}  = 0, \\
  \sigma_0^+ \sigma_1^- \ket{01}  = \ket{10},\ \
  \sigma_0^+ \sigma_1^- \ket{\mathrm{other\ than}\ 10}  = 0.
\end{align}
We can write $SE(\theta)$ satisfying Eq.~\ref{eq:SE} as
\begin{align}
  SE(\theta) = I \cos{(\theta/2)} - i K \sin{(\theta/2)} = e^{-i\frac{\theta}{2}K}.
\end{align}
The operator $K$ can be decomposed in terms of $X$ and $Y$ as
\begin{align}
  K = \frac{1}{2}\bigl(-X_0^{}Y_1^{}+Y_0^{}X_1^{}\bigr),
\end{align}
where all these two terms commute with each other. Therefore, $SE(\theta)$ can be given by two commuting two-qubit Pauli rotations.

\end{document}